\pdfoutput=1
\documentclass[%
 aip,
 amsmath,amssymb,
 reprint,%
]{revtex4-1}

\usepackage{graphicx}
\usepackage{dcolumn}
\usepackage{bm}
\usepackage[colorlinks,urlcolor=blue,linkcolor=blue,citecolor=blue]{hyperref}

\usepackage[utf8]{inputenc}
\usepackage[T1]{fontenc}
\usepackage{mathptmx}
\usepackage{etoolbox}
\usepackage{subfig}
\usepackage{xcolor}
\definecolor{FG}{HTML}{ed553b}

\makeatletter
\def\@email#1#2{%
 \endgroup
 \patchcmd{\titleblock@produce}
  {\frontmatter@RRAPformat}
  {\frontmatter@RRAPformat{\produce@RRAP{*#1\href{mailto:#2}{#2}}}\frontmatter@RRAPformat}
  {}{}
}%
\makeatother
\begin{document}

\preprint{AIP/123-QED}

\title[Ultra-Long-Range Bessel Beams via Leaky Waves with Mitigated Open Stopband]{Ultra-Long-Range Bessel Beams\\via Leaky Waves with Mitigated Open Stopband}
\author{E. Negri}
 \altaffiliation[Also at ]{Istituto per la Microelettronica e Microsistemi,\\Consiglio Nazionale delle Ricerche, 00133 Rome, Italy}
\affiliation{ 
Department of Information Engineering, Electronics and Telecommunications,\\Sapienza University of Rome, 00184 Rome, Italy
}
\email{edoardo.negri@uniroma1.it}

\author{F. Giusti}
\affiliation{Department of Information Engineering and Mathematics,\\University of Siena, 53100 Siena, Italy
}%

\author{W. Fuscaldo}%
\affiliation{%
Istituto per la Microelettronica e Microsistemi,\\Consiglio Nazionale delle Ricerche, 00133 Rome, Italy
}

\author{P. Burghignoli}
\affiliation{ 
Department of Information Engineering, Electronics and Telecommunications,\\Sapienza University of Rome, 00184 Rome, Italy
}

\author{E. Martini}
\affiliation{Department of Information Engineering and Mathematics,\\University of Siena, 53100 Siena, Italy
}%

\author{A. Galli}
\affiliation{ 
Department of Information Engineering, Electronics and Telecommunications,\\Sapienza University of Rome, 00184 Rome, Italy
}

\date{\today}

\begin{abstract}
Open stop-band (OSB) mitigation techniques are commonly used to improve the far-field radiating properties of leaky-wave antennas based on periodic structures. Recently, leaky waves have been proposed to focus energy in the near field through Bessel beams. However, the focusing character of Bessel beams is notably limited to a maximum distance known as the \emph{nondiffractive range}. In this work, an OSB mitigation technique is originally exploited to significantly extend the nondiffractive range of a Bessel beam generated by a leaky-wave launcher in the microwave/millimeter-wave range. A comprehensive analysis of this device is presented, comparing the performance of the proposed launcher with the typical structure of a leaky-wave Bessel-beam launcher where the OSB is not suppressed. Theoretical results corroborated by full-wave simulations demonstrate that the proposed device, designed around 30 GHz, achieves an impressive nondiffractive range of approximately 25\,m, which corresponds to 2500 vacuum wavelengths and to 50 times its aperture diameter of 50\,cm. These results look particularly attractive for, e.g., near-field communications and wireless power transfer applications, where focusing energy in narrow spots and over large distances is a key factor. 
\end{abstract}

\maketitle

Electromagnetic (EM) waves naturally spread their energy during propagation due to diffraction. However, in many practical scenarios such as secure communications \cite{bodet2024sub, reddy2023ultrabroadband}, radiative
near-field wireless power transmission \cite{Negri_TAP2023, heebl}, and imaging \cite{liu2020high}, a high-frequency focused EM beam is crucial to avoid transmitting power in undesired directions.
For this reason, Bessel beams (BBs) --- propagation-invariant cylindrical-wave solutions of the Helmholtz equation with focusing and self-healing features \cite{Durnin_nondiffracting_beams_scalar_theory, mcgloin} --- have gained significance. However, in practice, these intriguing properties are theoretically maintained only up to a finite distance, commonly known as \emph{nondiffractive range} \cite{Durnin_nondiffracting_beams_scalar_theory}. Therefore, it is essential to maximize this maximum covered distance across all frequency bands to fully harness the potential of BBs in real-world applications.

\begin{figure}[!t]
\centering
\includegraphics[width=\columnwidth]{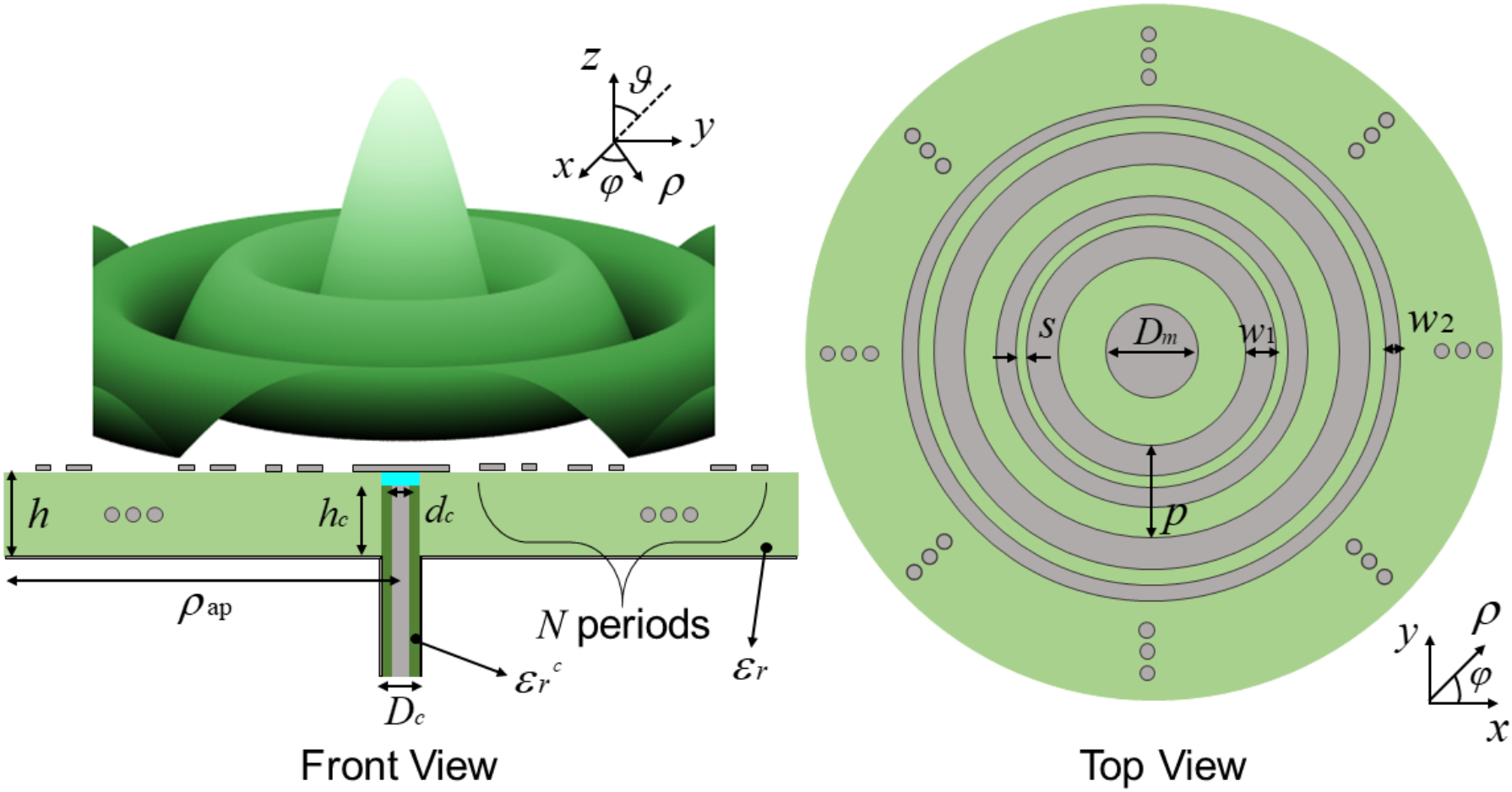}
\caption{Front and top view of the proposed wideband BB launcher based on a double asymmetric strip, with a pictorial representation of the excited $E_z$ field distribution. While different green areas represent the different dielectric permittivities of the coaxial cable and of the grounded dielectric slab, blue and grey colors are associated with vacuum and metal, respectively.} 
\label{fig:struct} 
\end{figure} 

\begin{figure*}[!t]
\centering
\subfloat[]{\includegraphics[width=0.333\textwidth]{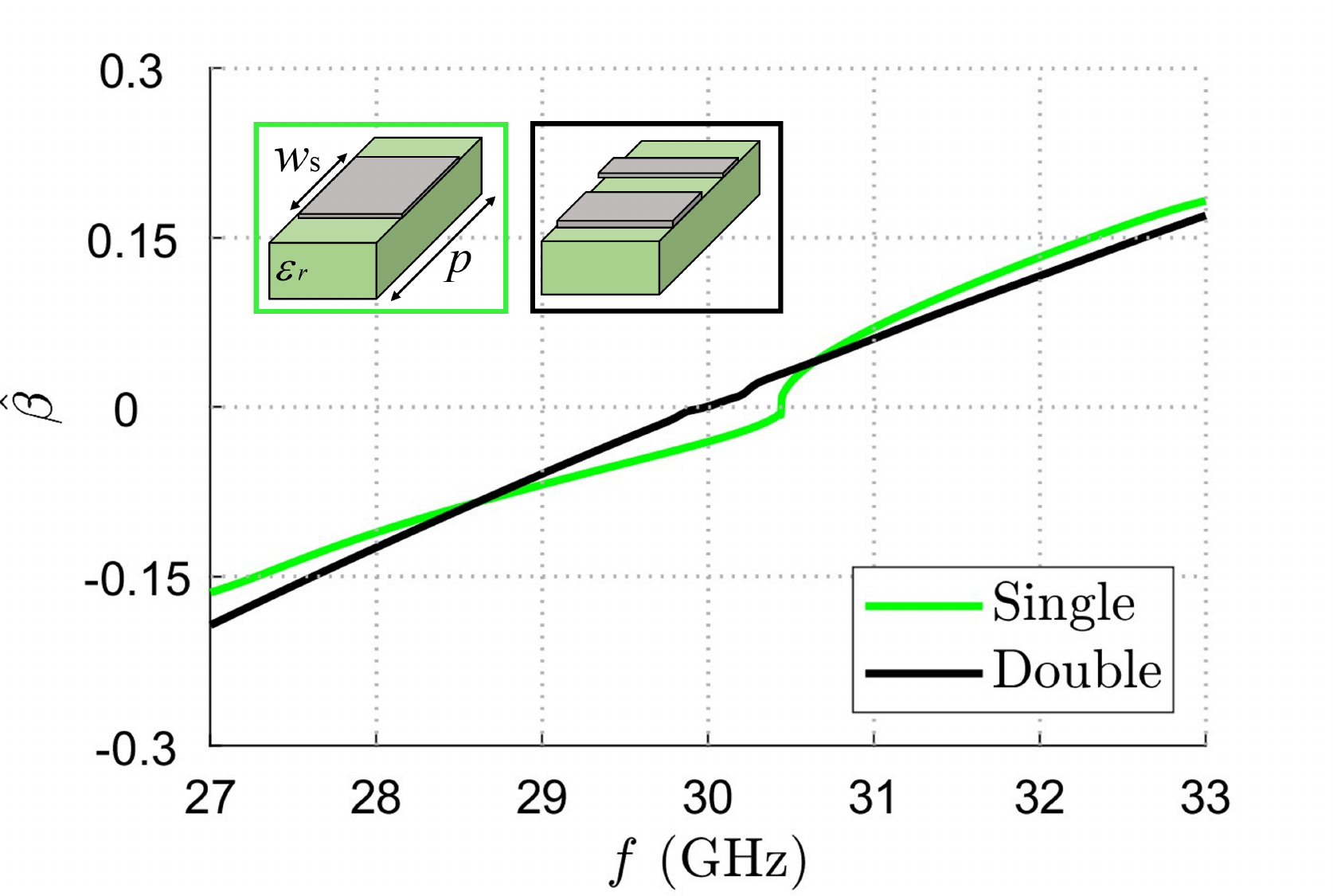}}\hfill
\subfloat[]{\includegraphics[width=0.333\textwidth]{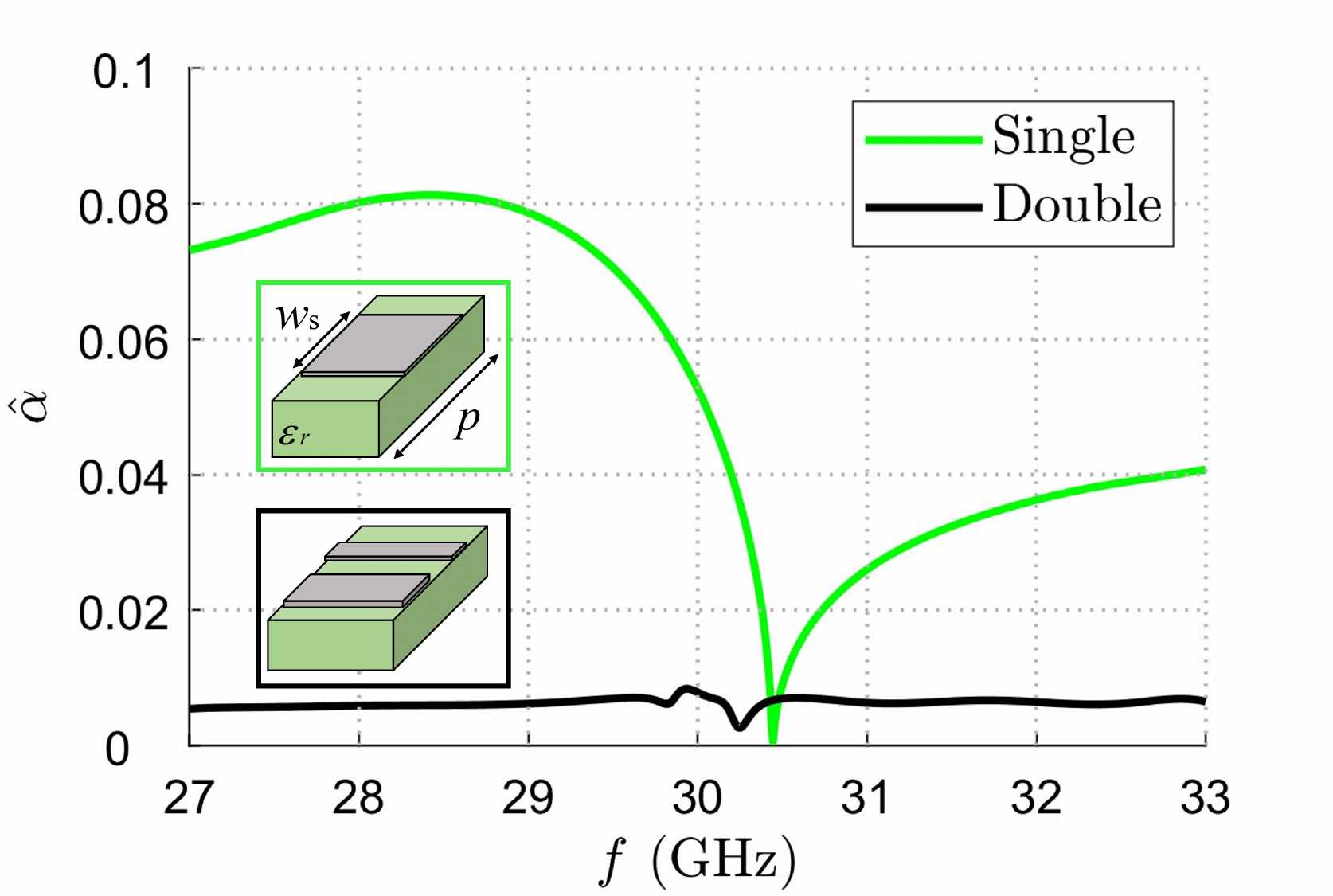}}\hfill
\subfloat[]{\includegraphics[width=0.333\textwidth]{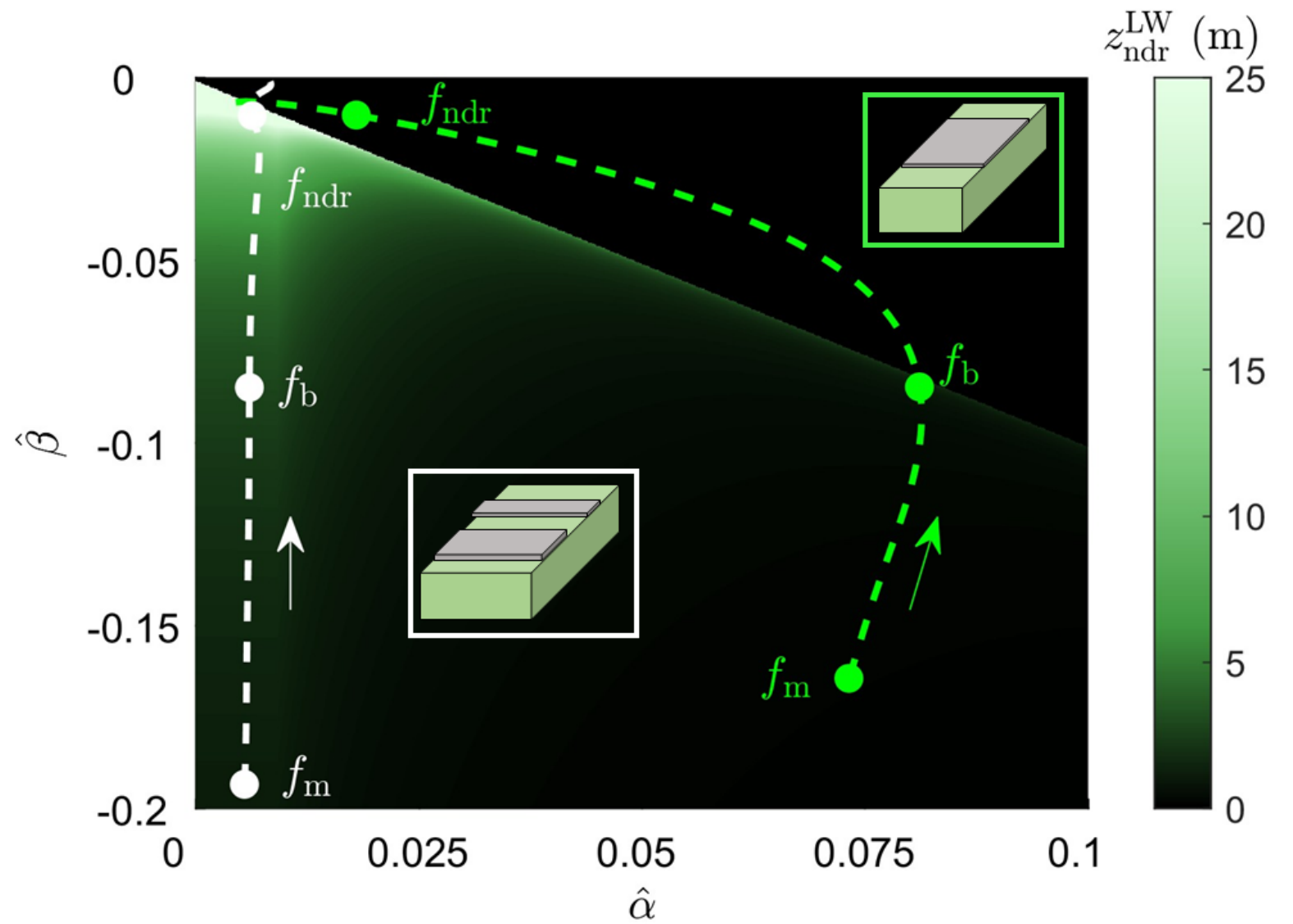}}
\caption{Bloch analysis of the single-strip symmetric (green curves) and double-strip asymmetric [black curves in (a) and (b)] unit cells. Leaky (a) phase, $\hat{\beta}$, and (b) attenuation, $\hat{\alpha}$, constants normalized with respect to the vacuum wavenumber $k_0$ are reported vs. frequency $f$. (c) Colormap of the leaky-wave nondiffractive range ($z_{\mathrm{ndr}}^{\mathrm{LW}}$) as a function of the normalized leaky phase, $\hat{\beta}$, and attenuation, $\hat{\alpha}$, constants. The white and green dashed lines represent the synthesized $\hat{\beta}$ and $\hat{\alpha}$ values as the frequency varies ($f$ increases in the arrow direction) for the double- and single-strip unit-cell configurations, respectively. Three interesting frequency points $f_\mathrm{m}$, $f_\mathrm{b}$ and $f_\mathrm{ndr}$ are here marked. As a reference, the unit-cell representation of each considered case is linked to the corresponding curve in a color-code manner.} 
\label{fig:disp} 
\end{figure*}

Although BBs have been preliminarily studied in optics since the `80s \cite{Durnin_nondiffracting_beams_scalar_theory}, the growing need in focusing EM energy at lower frequencies has renewed research in this area. As a result, the last decade has seen an exponential growth of works exploring  both theoretical and experimental aspects of BBs in
the microwave/millimeter-wave domain \cite{ettorre_chapter}. In these frequency ranges, a planar, cost-effective, single-feeder, and easy-to-fabricate solution is given by \emph{leaky-wave} BB launchers. In particular, \emph{wideband} BB launchers \cite{comite_radial} are preferable with respect to their resonant counterparts \cite{Negri_Micromachines} for achieving a higher nondiffractive range $z_{\rm ndr}$, due to their inherently larger radiating aperture. In fact, the value of $z_{\rm ndr}$ scales linearly with the aperture radius $\rho_{\rm ap}$. By exploiting a ray-optics approximation \cite{Durnin_nondiffracting_beams_scalar_theory}, these two quantities are more precisely connected by the so-called \emph{axicon angle} $\theta_0$ through the equation: $z_{\rm ndr}=\rho_{\rm ap}\cot\theta_0$.

The axicon angle represents the pointing direction of the rays emerging from the BB aperture source with respect to the vertical $z$ axis (see Fig.\,\ref{fig:struct}). The closer the rays align with the broadside direction, implying that the axicon angle approaches zero, the greater the increase in the nondiffractive range. As reported in the literature \cite{comite_radial,Negri_Micromachines}, the $\theta_0$ value in leaky-wave launchers is related to the complex radial wavenumber $k_\rho=\beta-j\alpha$, with $\beta$ and $\alpha$ being the so-called leaky phase and attenuation constants, respectively. Since one typically has that $\theta_0 = \arcsin{(\beta/k_0)}$ (with $k_0$ being the vacuum wavenumber), the nondiffractive range can be enhanced by reducing the radial phase constant as much as possible, while keeping the $\alpha$ value sufficiently small, but not vanishing \cite{fuscaldo_BGB, Fuscaldo_Focusing_LW, Negri_ISAP}.

However, in wideband BB launchers, when $\beta$ tends to zero, the value of $\alpha$ changes rapidly: first, it  exhibits a highly peaked behavior due to an accumulation of reactive energy, and then drops to zero at the broadside frequency \cite{antennaTheory}. This behavior is attributed to the \emph{open-stopband} (OSB) phenomenon, which rises from the coupling between a pair of oppositely directed spatial harmonics. Different techniques were investigated to mitigate or suppress the OSB, thereby improving the \emph{far-field} radiating properties of 1-D or 2-D leaky-wave antennas \cite{Burghignoli,Al_Bassam_OSBsuppression,Liu18, GiustiOSB}. In particular, a longitudinal asymmetry design principle was applied \cite{Liu18} using two similar but \emph{unequal} discontinuities inside the unit cell\cite{Comite2019}.

In this work, an OSB mitigation technique is originally exploited to improve the \emph{near-field} properties of wideband BB launchers. In particular, a radial unit cell with a \emph{double asymmetric discontinuity} has been considered to extend as much as possible the nondiffractive range of a BB. In order to prove the effectiveness of the proposed design, we compare this structure with a typical wideband BB launcher \cite{comite_radial} based on a \emph{single discontinuity} where the OSB is present. A comprehensive Bloch analysis of the two unit cells is performed using the transfer-matrix method to compute the real and imaginary parts of the relevant leaky wavenumber \cite{CascadedTMatrixApproach,Mesa_MMTMM,giustiMMTMA}. It is shown that the BB launcher based on the double-discontinuity configuration is able to generate a BB distribution over an impressively long range. With an aperture radius of 24.18\,cm, the maximum nondiffractive distance around 30\,GHz reaches nearly 25\,m, namely, 50 times the aperture diameter and $2500\lambda_0$, with $\lambda_0=1$\,cm being the vacuum wavelength at the central frequency $f_0=30$\,GHz. The leaky-wave theoretical analysis is compared with full-wave simulations, showing a remarkable agreement between the two approaches.

The analysis starts with the design of the wideband BB launchers and the evaluation of their performance through the leaky-wave theory \cite{Fuscaldo_Focusing_LW}. The structures considered in this work are radially periodic leaky-wave radiators constituted by a grounded dielectric slab (with relative permittivity $\varepsilon_\mathrm{r}=3.02$, negligible losses, and thickness \mbox{$h=1.4$\,mm)} with an annular metal strip grating on top \cite{comite_radial}.
As a consequence, the device performance is strictly related to the leaky phase and attenuation constants \cite{Fuscaldo_Focusing_LW}. The main innovative contribution of this work is the use of an asymmetric double-strip unit cell able to mitigate the OSB \cite{Comite2019} in order to enhance the $z_{\rm ndr}$ value: the OSB mitigation is indeed a crucial feature for the generation of long-nondiffractive-range BB since it allows us to achieve a non-vanishing, sufficiently small, leakage constant $\alpha$ as the leaky phase constant $\beta\to 0$.

\begin{figure*}[!t]
\centering
\subfloat[]{\includegraphics[width=0.24\textwidth]{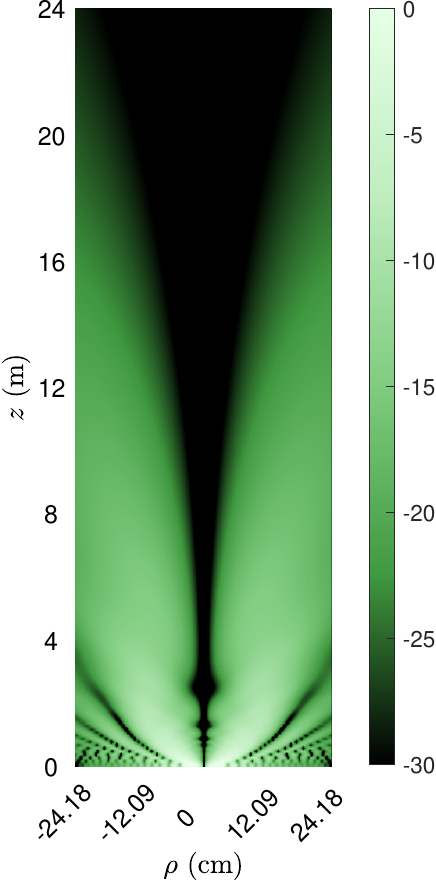}}
\hfill
\subfloat[]{\includegraphics[width=0.24\textwidth]{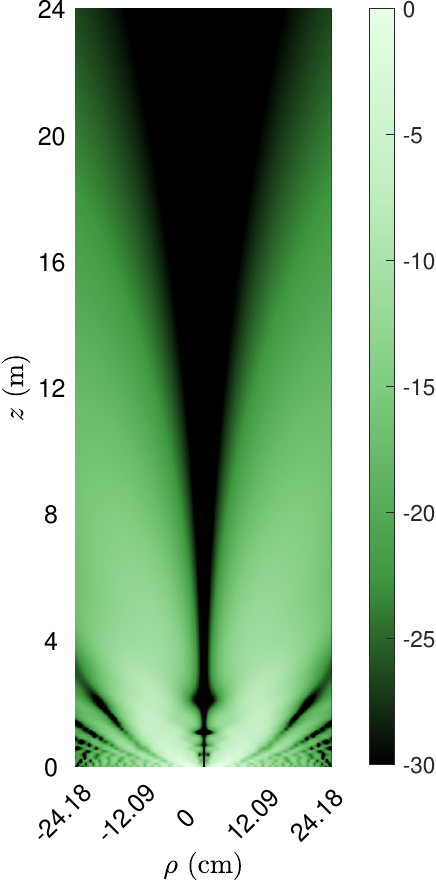}} \hfill
\subfloat[]{\includegraphics[width=0.24\textwidth]{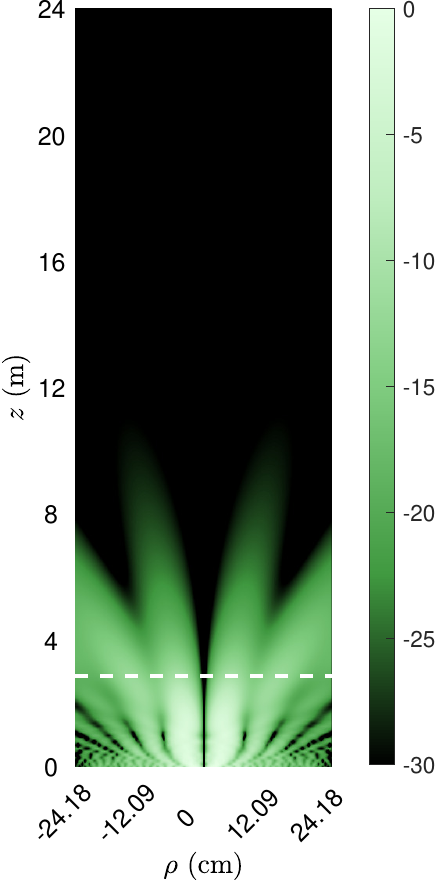}} \hfill
\subfloat[]{\includegraphics[width=0.24\textwidth]{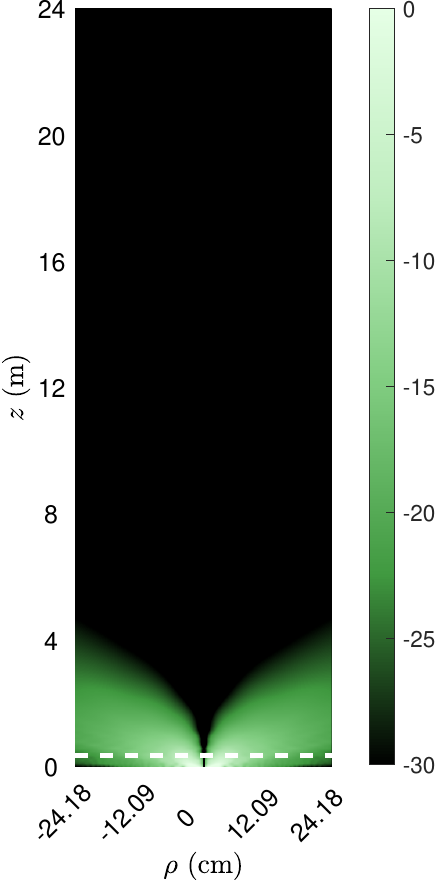}} \hfill
\caption{Colormap in dB of the electric-field radial component --- normalized with respect to its maximum --- obtained through the radiation integral of the (a) leaky-wave and (b) full-wave aperture-field distributions for the double-strip unit-cell configuration at $f=f_{\rm ndr}$. The full-wave evaluation of the same field component is shown for the case of a (c) double- and (d) single-strip unit cell at $f=f_{\rm b}$ (the leaky-wave nondiffractive range is represented through a white dashed line).} 
\label{fig:field} 
\end{figure*}

In  order to assess the effectiveness of the OSB mitigation in terms of nondiffractive-range enhancement, the Bloch analysis of the constituent unit cell is performed using the transfer-matrix method \cite{CascadedTMatrixApproach,Mesa_MMTMM,giustiMMTMA}. To this end, 
the linearized 1-D counterpart of the structure has been considered \cite{comite_radial, podilchak2014analysis}.  Two different unit cells are analyzed: the first one has two asymmetric strips of width $w_1=1.07$\,mm and $w_2=0.86$\,mm separated by a distance $s=2$\,mm (see Fig.\,\ref{fig:struct}), while the second one has a single strip of width $w_s=w_1+w_2$ (see Fig.\,\ref{fig:disp}). In both cases the period is equal to $p=7.8$\,mm.

For this analysis, two waveguide ports with a height of $5.6\lambda_0$ are defined along the periodicity direction. At the two ports, only the propagation of the fundamental TM$_0$ surface-wave mode of the grounded dielectric slab is considered. Perfect magnetic boundaries and an open boundary condition are assigned along $y$ and $z$ (see insets in Fig.~\ref{fig:disp}), respectively.
For both the cases, the transfer matrix of the single unit cell is extracted with CST full-wave simulations by multiplying the transfer matrix obtained simulating $N=8$ unit cells with the inverse of the one of $N=7$ unit cells \cite{CascadedTMatrixApproach}. Additional unit cells in the CST simulation only introduce  numerical noise, thus negatively affecting the accuracy of the results. 

The results of the Bloch analysis of the single-strip symmetric unit cell are reported with green curves in Fig.\,\ref{fig:disp}. For convenience, both phase and attenuation constants are normalized with respect to the vacuum wavenumber $k_0$ and indicated as $\hat{\beta}=\beta/k_0$ and $\hat{\alpha}=\alpha/k_0$, respectively. The previously described OSB effect can be clearly seen both in the perturbation of $\hat{\beta}$ and in the null of $\hat{\alpha}$ at the broadside frequency of 30.4\,GHz. In contrast, the dispersion curves for the proposed double-strip asymmetric unit cell, shown as black curves in the same figure for comparison, exhibit a mitigated OSB effect, evident as a minor perturbation in both $\hat{\beta}$ and $\hat{\alpha}$ near the broadside frequency. 

Once the dispersion curves of the leaky phase and attenuation constants are retrieved, it is possible to theoretically evaluate the performance of the proposed BB launcher. In uniform radiating apertures, the BB nondiffractive range is given by the well-known ray-optics approximation ${z_{\rm ndr}=\rho_{\rm ap}\cot\theta_0}$ \cite{Durnin_diffraction_free_beams}. However, the effective nondiffractive range of a leaky-wave BB slightly differs from the ideal one since it is affected by the attenuated aperture-field profile. In particular, by indicating the normalized aperture radius as $\Bar{\rho}_{\rm ap}=\rho_{\rm ap}/\lambda_0$, it reads \cite{fuscaldo_BGB}:
\begin{equation}
z_{\rm{ndr}}^{\rm LW} = \begin{cases}
z_{\rm ndr} \frac{\ln\sqrt{2}}{\pi\hat{\alpha}\Bar{\rho}_{\text {ap}}}, & \frac{\ln\sqrt{2}}{\pi\Bar{\rho}_{\text {ap}}}<\hat{\alpha} \ll 1\;,\\
z_{\rm ndr}, & \hat{\alpha} \leq \frac{\ln\sqrt{2}}{\pi\Bar{\rho}_{\text {ap}}} \ll 1\;.
\end{cases}
\label{eq:z_ndr}
\end{equation}
It is worth noting that, while the upper limit $\hat{\alpha}\ll 1$ is required for a physically meaningful leaky wave, the switching condition between the two $z_{\rm ndr}^{\rm LW}$ definitions, $\hat{\alpha}=\ln\sqrt{2}/(\pi\Bar{\rho}_{\text {ap}})$, is imposed to avoid nonphysical cases for which $z_{\rm ndr}^{\rm LW}>z_{\rm ndr}$ \cite{fuscaldo_BGB}.

The axicon angle $\theta_0$ is commonly evaluated in leaky-wave BB launchers through the normalized leaky phase constant by exploiting the simple ray-optics approximation: \mbox{$\theta_0=\arcsin\hat{\beta}$} \cite{comite_radial, Fuscaldo_Focusing_LW}. However, in this work, since $\hat{\beta}$ tends to zero for maximizing $z_{\rm ndr}$, the contribution of the normalized leaky attenuation constant is nonnegligible and it has to be considered for evaluating the leaky-wave pointing angle through \cite{Lovat_Broadside_LW}:
\begin{equation}
    \theta_0 = \arcsin{\sqrt{\hat{\beta}^2-\hat{\alpha}^2}}\;.
    \label{eq:theta0}
\end{equation}%

At this point, given \mbox{$\rho_{\rm ap}=24.18$\,cm}, the nondiffractive range $z_{\rm ndr}^{\rm LW}$ of a generic leaky-wave BB launcher can be computed through \eqref{eq:z_ndr} with the $z_{\rm ndr}$ evaluation through the definition of the axicon angle in \eqref{eq:theta0}. As shown in Fig.~\ref{fig:disp}(c), the nondiffractive range reaches its maximum value of about 25\,m  (approximately 50 times the aperture diameter and 2500$\lambda_0$) when both $\hat{\beta}$ and $\hat{\alpha}$ are small and nonvanishing. It is worthwhile to point out that, when $|\hat{\beta}|<\hat{\alpha}$ [see black region in Fig.~\ref{fig:disp}(c)], we are working in the \emph{reactive regime} of the propagating leaky wave (see \cite{fuscaldo_MTT} for further details) and, thus, we do not consider the possibility to properly generate a BB.

The considered single- and double-strip unit-cell configurations are able to excite a leaky mode whose frequency-dispersive behavior is represented by the green and white curves shown in Fig.~\ref{fig:disp}(c), respectively. These latter are displayed starting from the minimum considered frequency $f_{\rm m} = 27$\,GHz in the region where an inward cylindrical leaky wave is achieved ($\hat{\beta}<0$ and $\hat{\alpha}>0$). From a closer inspection of Fig.~\ref{fig:disp}(c), the importance of the OSB suppression is clear: the double-strip unit-cell configuration admits a leaky wavenumber with simultaneously low and flat values for $\hat{\beta}$ and $\hat{\alpha}$ and, in turn, very high nondiffractive-range values. Conversely, the single-strip unit cell, due to the OSB effect, shows large and rapidly changing $\hat{\alpha}$ values when $\hat{\beta}\to0$, thus preventing the generation of a BB over a wide spatial region (the green dashed curve mainly falls in the black, reactive region with $|\hat{\beta}|<\hat{\alpha}$). In this case, the best working condition is at the frequency $f_{\rm b}=28.6$\,GHz, where $|\hat{\beta}|\simeq\hat{\alpha}$, with $\hat{\beta}=-0.084$ and $\hat{\alpha}=0.081$, corresponding to $z_{\rm ndr}^{\rm LW}=0.34$\,m. The same $\hat{\beta}$ is obtained at the same working frequency for the double-strip case with a much lower attenuation constant ($\hat{\alpha}=0.006$) thanks to the OSB mitigation; in this manner, a $z_{\rm ndr}^{\rm LW}=z_{\rm ndr}=2.87$\,m is achieved, which clearly shows the detrimental effect of large $\hat{\alpha}$ values in the generation of long nondiffractive ranges. 

Another interesting working point is the frequency \mbox{$f_{\rm ndr}=29.79$\,GHz} (where $\hat{\beta}=-0.01$ and $\hat{\alpha}=0.006$, for the double-strip unit cell), for which the maximum nondiffractive range of $z_{\rm ndr}=24.18$\,m is obtained by the proposed wideband BB launcher, a performance that would be unattainable without mitigating the OSB. The condition for which the same $\hat{\beta}$ is synthesized by the single-strip unit cell falls in the reactive region (viz., $|\hat{\beta}|<\hat{\alpha}$) at $f_{\rm ndr}=30.39$\,GHz with $\hat{\alpha}=0.018$.

\begin{figure}[!t]
\centering
\includegraphics[width=0.8\columnwidth]{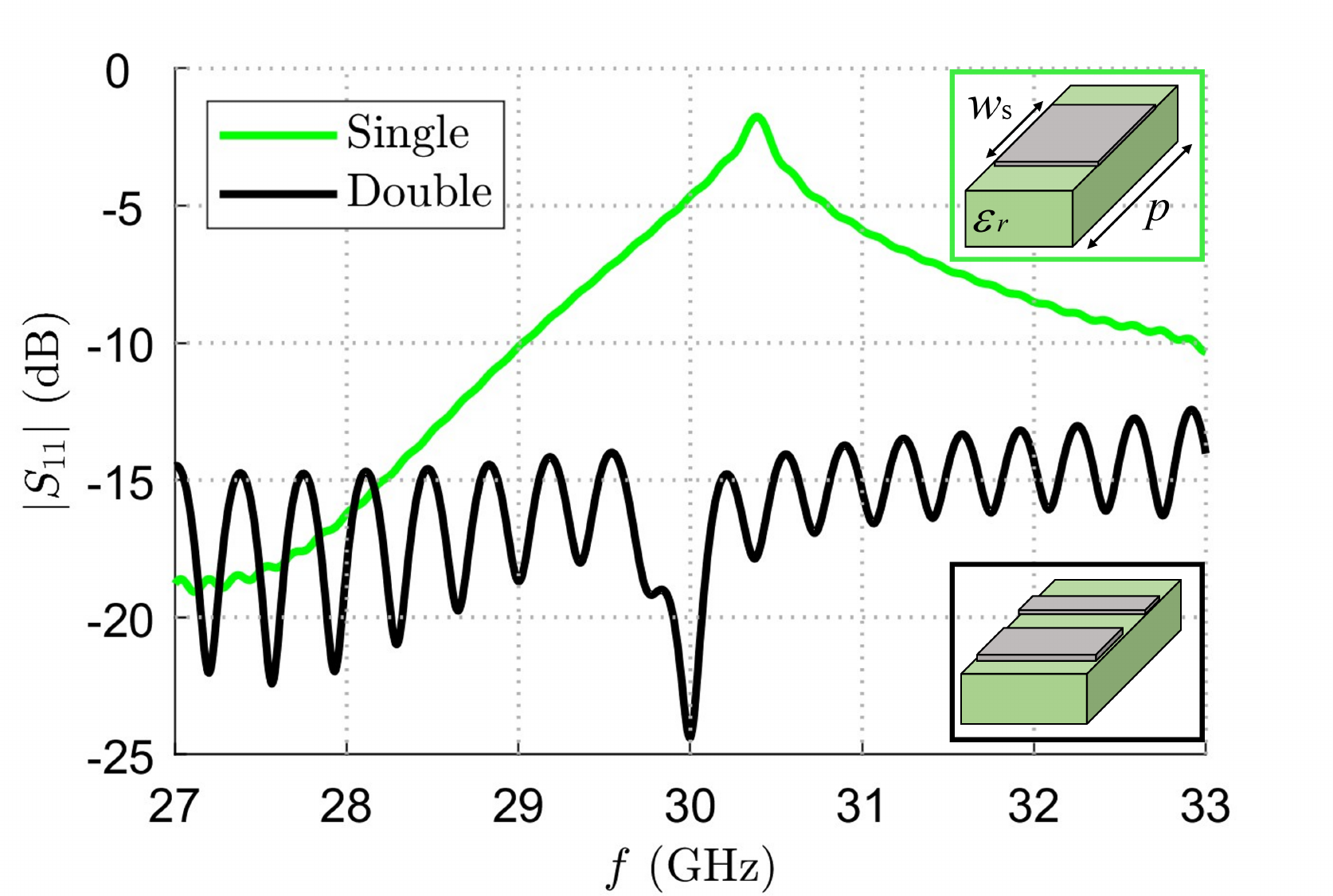}
\caption{$|S_{11}(\rm{dB})|$ parameter vs. frequency $f$ of the proposed long-nondiffractive-range double-strip BB launcher (black solid line --- inset on the bottom right corner) and the conventional single-strip BB launcher (green solid line --- inset on the top right corner).} 
\label{fig:S11} 
\end{figure} 

In order to corroborate the abovementioned generation of a BB over a wide nondiffractive region, the near-field evaluation of the device has to be addressed. The direct derivation of the electromagnetic field distribution from the full-wave solver is however computationally expensive due to the very large extension of the spatial domain. For this reason, the simulation of a 3-D model of the device has been applied to compute the full-wave \emph{aperture field distribution}. As well known, once the tangential aperture field distribution is derived, the equivalence theorem can be applied to find the equivalent electric and magnetic surface currents on the aperture plane \cite{Bal_Adv}. At that point, the near-field distribution is computed through the Huygens--Fresnel radiation integral (see, e.g., \cite{Fuscaldo_Focusing_LW} for the field-evaluation procedure).

For the proper generation of the field in our wideband BB launcher a very simple coaxial-cable feeder, penetrating the ground plane of the structure, can be chosen. This source is able to excite a transverse-magnetic (TM) polarized field. While the dielectric permittivity $\varepsilon_\mathrm{r}^{c}=2.1$ and the dimensions of the coaxial cable conductors \mbox{($D_c=0.99$\,mm and $d_c=0.31$\,mm --- see Fig.\,\ref{fig:struct})} are standard, the diameter of a metallic matching disk \mbox{$D_m=5.38$\,mm} and the penetration level of the coaxial cable \mbox{$h_c=1.32$\,mm} in the grounded dielectric slab (drilled at the center with a diameter $D_c$ --- see Fig.\,\ref{fig:struct}) are tuned for matching purposes. The $S_{11}$ obtained through a full-wave simulation of the double-strip configuration is illustrated in Fig.~\ref{fig:S11}, showing a very good matching over the whole frequency band. By using the same feeding scheme for the single-strip configuration of the wideband BB launcher, a strong mismatch appears around 30.4\,GHz: this is due to the presence of the OSB \cite{GiustiOSB}. It is therefore clear that the OSB mitigation is important not only to simultaneously have a small and nonvanishing value of $\hat{\alpha}$ as $\hat{\beta}\to 0$, but also to avoid any matching issue.

As shown in previous works \cite{Ettorre_BB, Ettorre_BB_exp}, a VED is an accurate model of the abovementioned coaxial feeders, thus the total aperture field distribution is suitably approximated by the azimuthally symmetric, TM-polarized, leaky-wave contribution \cite{Ip_LW}. By considering the leaky-wave aperture field for wideband BB launchers \cite{comite_radial} with the leaky phase and attenuation constants in Fig.~\ref{fig:disp}, the theoretical near-field distribution can be computed through the equivalence principle and the Huygens--Fresnel radiation integral \cite{fuscaldo_BGB, Bal_Adv}.

The results reported in Fig.\,\ref{fig:field}(a) and (b) are the near-field distributions of the proposed double-strip wideband BB launcher at $f=f_{\rm ndr}$, obtained through the radiation integral of the leaky-wave and full-wave aperture field profiles, respectively. The excellent agreement corroborates both the generation of the abovementioned BB with $z_{\rm ndr}\simeq24$\,m through an aperture radius $\rho_{\rm ap}=24.18$\,cm and the validity of the leaky-wave analysis through a full-wave simulation. Figure \ref{fig:field}(c) shows the simulated field distribution for the launcher with double-strip unit cells at $f_{\rm b}=28.62$\,GHz, where $\hat{\beta}=-0.084$ and $\hat{\alpha}=0.006$. As discussed above, this phase-constant value corresponds to the theoretically ``best'' working condition for the single-strip unit-cell configuration. Therefore, in Fig.~\ref{fig:field}(d), the BB generated at $f=f_{\mathrm b}$ by the launcher with the typical single-strip unit cell is reported for comparison. As expected, a leaky-wave nondiffractive range of only 34\,cm (white dashed line in Fig.\,\ref{fig:field}(d)) is reached in this case. With the same $\hat{\beta}$, a nondiffractive range $z_{\rm ndr}=2.87$\,m [see white dashed line in Fig.\,\ref{fig:field}(c)] is achieved thanks to the OSB mitigation provided by the double-strip configuration. Therefore, the importance of the OSB suppression in leaky-wave BB launcher is again clearly visible by a direct comparison of the electric-field distributions in Figs.\,\ref{fig:field}(c) and (d).

In conclusion, this work has demonstrated the critical role of open-stopband mitigation in extending the nondiffractive range of leaky-wave wideband Bessel-beam launchers. In particular, this objective has been addressed through the design of an asymmetric double-strip radial unit cell for the annular metal strip grating constituting the radiating aperture of the device, given by a simple grounded dielectric substrate fed by a coaxial probe. The theoretical leaky-wave design remarkably agrees with full-wave simulations, confirming the possibility of generating a Bessel beam over extremely long distances. In particular, with an aperture radius of about 25\,cm, a nondiffractive range of approximately 25\,m has been achieved --- equivalent to 50 times the aperture diameter and 2500 vacuum wavelengths. These findings highlight the potential of Bessel beams for innovative and practical millimeter-wave applications, setting the stage for new advancements in the area of long-distance energy focusing.

\begin{acknowledgments}
E. N. and W. F. acknowledge the project PRIN 2022 “SAFE” (Spiral and Focused Electromagnetic fields) 2022ESAC3K, Italian Ministry of University and Research (MUR), financed by the European Union, Next Generation EU. A. G. thanks for the support of the European Union under the Italian National Recovery and Resilience Plan (NRRP) of NextGenerationEU, partnership on “Telecommunications of the Future” (PE00000001 - program “RESTART”).
\end{acknowledgments}

\section*{Data Availability Statement}

The data that support the findings of this study are available from the corresponding author upon reasonable request.


%

\end{document}